%% file: main.tex
\documentclass{article}
\input{packages}

\input{macros}

\title{HEMPPCAT: MIXTURES OF PROBABILISTIC PRINCIPAL COMPONENT ANALYSERS FOR DATA WITH HETEROSCEDASTIC NOISE}
\name{Alec S. Xu, Laura Balzano, Jeffrey A. Fessler%
\thanks{Supported in part by NSF Grant IIS 1838179.}}
\address{EECS Department, University of Michigan-Ann Arbor
}
\begin{document}
%
\maketitle
\input{body}
\vfill\pagebreak



\bibliographystyle{IEEEbib}
\bibliography{strings,refs,master}

\include{supp} 

\end{document}

%% file: packages.tex
\usepackage{spconf}
\usepackage{outline}
\usepackage{pmgraph}
\usepackage[normalem]{ulem}
\usepackage[utf8]{inputenc}
\usepackage{amssymb}
\usepackage{hyperref}
\usepackage{amsmath}
\usepackage{graphicx}
\usepackage[text={6.5in,9in},centering]{geometry}
\usepackage{times}
\usepackage{xcolor}
\usepackage{xspace}
\usepackage[colorinlistoftodos]{todonotes} 
\usepackage{bm} 
\usepackage{subcaption}
\usepackage{adjustbox}

\usepackage{epstopdf}
\AppendGraphicsExtensions{.tiff}
\graphicspath{{fig/}} 

\usepackage{epsfig}
\usepackage{tikz}
\usepackage{algpseudocode}
\usepackage{algorithm}
\usepackage{mathrsfs}

%% file: macros.tex
\long\def\comment#1{}
\newcommand{\Frac}[2]{#1/#2}
\long\def\red#1{\bgroup\color{red}#1\egroup}
\newcommand{\vsp}[1]{\vspace*{-#1em}}
\newcommand{\normF}[1] {\| #1 \|_{\mathrm{F}}}

\newcommand{\iid}{IID\xspace}

\def\QED{~\rule[-1pt]{5pt}{5pt}\par\medskip}

\newcommand{\argmax}{\operatorname{argmax}}

\newcommand{\xmath}[1] {\ensuremath{#1}\xspace}
\newcommand{\blmath}[1] {\xmath{\bm{#1}}}
\newcommand{\A}{\blmath{A}}
\newcommand{\B}{\blmath{B}}
\newcommand{\C}{\blmath{C}}

\newcommand{\F}{\blmath{F}}
\newcommand{\K}{\blmath{K}}
\newcommand{\I}{\blmath{I}}

\newcommand{\U}{\blmath{U}}

\newcommand{\X}{\blmath{X}}

\newcommand{\x}{\blmath{x}}
\newcommand{\y}{\blmath{y}}
\newcommand{\z}{\blmath{z}}
 
\newcommand{\0}{\blmath{0}} 
\renewcommand{\Re}{\mathbb{R}}

\newcommand{\bmu}{\blmath{\mu}}
\newcommand{\btheta}{\blmath{\theta}}
\newcommand{\beps}{\blmath{\epsilon}}

\newcommand{\Nor}{\mathcal{N}} 
\newcommand{\Com}{\mathbb{C}}
\newcommand{\LL}{\mathcal{L}} 
\newcommand{\var}{\blmath{v}} 

\newcommand{\exprm}{\mathrm{exp}} 
\newcommand{\lnrm}{\mathrm{ln}} 
\newcommand{\trrm}{\mathrm{tr}} 
\newcommand{\detrm}{\mathrm{det}}

\newcommand{\yli}{\y_{\ell, i}}
\newcommand{\yblij}{\bar{\y}_{\ell ij}}

\newcommand{\glij}{g_{\ell ij}}
\newcommand{\Rlijt}{R_{\ell ij}^{(t)}}

\newcommand{\zl}{\z_{\ell, i}}
\newcommand{\zlj}{\xmath{\z_{\ell ij}}} 
\newcommand{\zle}{\langle \zlj \rangle}
\newcommand{\zlC}{\langle \zlj \zlj^T \rangle}

\newcommand{\Ft}{\F^{(t)}} 
\newcommand{\Fjt}{\F^{(t)}_j} 
\newcommand{\FjT}{\F^T_j} 
\newcommand{\FjtT}{\F^{(t) \: T}_j} 
\newcommand{\Fjtnew}{\F^{(t+1)}_j}

\newcommand{\vl}{v_{\ell}}
\newcommand{\vart}{\var^{(t)}} 
\newcommand{\varlt}{v^{(t)}_\ell} 
\newcommand{\varltnew}{v^{(t+1)}_\ell} 

\newcommand{\pitnew}{\pi_j^{(t+1)}}
\newcommand{\M}{\blmath{M}}
\newcommand{\p}{\blmath{p}}

\newcommand{\bmujtnew}{\bmu_j^{(t+1)}}

\newcommand{\Clj}{\C_{\ell, j}}
\newcommand{\Cljinv}{\Clj^{-1}}
\newcommand{\Mlj}{\M_{\ell, j}}
\newcommand{\Mljinv}{\M_{\ell, j}^{-1}}
\newcommand{\Mljt}{\M_{\ell, j}^{(t)}}
\newcommand{\Mljtinv}{\M_{\ell, j}^{(t) \: -1}}

\newcommand{\sumj}{\sum_{j=1}^{J}}
\newcommand{\suml}{\sum_{\ell=1}^{L}}
\newcommand{\sumi}{\sum_{i=1}^{n_\ell}}
\newcommand{\sumlimj}{\sum\limits_{j=1}^{J}}
\newcommand{\sumliml}{\sum\limits_{\ell=1}^{L}}
\newcommand{\sumlimi}{\sum\limits_{i=1}^{n_\ell}}

\newcommand{\sumlij}{\suml \sumi \sumj}

\newcommand{\expec}{\mathbb{E}}

%% file: body.tex
\begin{abstract}
Mixtures of probabilistic principal component analysis (MPPCA)
is a well-known mixture model extension of principal component analysis (PCA).
Similar to PCA, MPPCA assumes the data samples in each mixture contain homoscedastic noise.
However, datasets with heterogeneous noise across samples are becoming increasingly common,
as larger datasets are generated by collecting samples from several sources with varying noise profiles.
The performance of MPPCA is suboptimal for data with heteroscedastic noise across samples.
This paper proposes a heteroscedastic mixtures of probabilistic PCA technique (HeMPPCAT)
that uses
a generalized expectation-maximization (GEM) algorithm
to jointly estimate the unknown underlying factors, means, and noise variances
under a heteroscedastic noise setting. 
Simulation results 
illustrate the improved factor estimates
and clustering accuracies
of HeMPPCAT
compared to MPPCA.
\end{abstract}

\vsp{1}
\section{Introduction}
\label{sec:intro}
\vsp{1}

PCA
is a well-known unsupervised dimensionality reduction method for high-dimensional data analysis. 
It has been extended to capture a mixture
of low-dimensional affine subspaces.
When this mixture model is derived through a probabilistic perspective,
it is called Mixtures of Probabilistic PCA (MPPCA) \cite{tipping:99:mop}.
MPPCA models
are a statistical extension
of union-of-subspace models
\cite{lu:08:ssf,blumensath:11:sar,eldar:09:rro}
and are also related to subspace clustering methods
\cite{vidal:05:gpc,vidal:11:sc}. 

One can apply MPPCA
to many engineering and machine learning tasks,
such as image compression and handwritten digit classification.

However, a limitation 
of PPCA and MPPCA is that
they both model the noise
as independent and identically distributed (\iid) with a shared
variance, i.e., homoscedastic.
Consequently, the performance of PPCA and MPPCA
can be suboptimal 
for heteroscedastic noise conditions.
Heterogeneous datasets are increasingly common,
e.g.,
when combining samples 
from several sources 
\cite{cabani2021maskedface}, or
samples collected under
varying ambient conditions 
\cite{georghiades2001few}.
Recently an algorithm called
HePPCAT was developed
to extend PPCA
to data with
heteroscedastic noise across samples
\cite{hong:21:hpp},
but no corresponding methods exist
for a \emph{mixture} of PPCA models.
This paper generalizes MPPCA
to introduce a \textbf{He}teroscedastic \textbf{MPPCA} \textbf{T}echnique (HeMPPCAT):
an MPPCA method for data with heteroscedastic noise.
This paper presents the statistical data model,
a GEM algorithm,
and results with both synthetic data
and motion segmentation
from real video data.

\vsp{1}

\section{Related Work}
\label{sec:related}

\vsp{1}

\subsection{MPPCA}
\vsp{1}

The original MPPCA approach
\cite{tipping:99:mop}
models $n$ data samples in $\Re^d$
as arising from $J$ affine subspaces:
\begin{equation} \label{eq:mppca_model}
    \y_i = \F_j\z_i + \bmu_j + \beps_{i}
\end{equation}
for all $i \in \{1, \dots, n\}$
and some class index
$j = j_i \in \{1, \dots, J\}$.
Here
$\F_1, \dots, \F_J \in \Re^{d \times k}$ are deterministic factor matrices to estimate,
$\z_i \sim \Nor(\0_k, \I_k)$ are \iid factor coefficients,
$\bmu_1, \dots, \bmu_J \in \Re^d$ are unknown deterministic mean vectors to estimate,
$\beps_{i} \sim \Nor(\0_d, v_j \I_d)$ are \iid noise vectors,
and $v_1, \dots, v_J$ are unknown noise variances to estimate.

MPPCA assumes all samples from mixture component $j$ have the same noise variance $v_j$.
In contrast,
our proposed
HeMPPCAT allows each sample from mixture $j$
to come from one of $L \leq n$ noise groups,
where $L \neq J$ in general.

\vsp{1}
\subsection{Probabilistic PCA for Heteroscedastic Signals}
\vsp{1}

The authors of
\cite{collas:21:ppf}
developed a Riemannian optimization method to perform an MPPCA-like method on signals with heterogeneous power levels. They model $n$ data samples in $\Com^d$ as 
\begin{equation}
\label{eq:collas}
    \y_i = \sqrt{\tau_j}\U_j \z_i + \beps_i
\end{equation}
for all $i = \{1, \dots, n\}$ and some $j = j_i \in \{1, \dots, J\}$.
The signal powers $\tau_1, \dots, \tau_J \in \Re^+$,
known as \textit{signal textures}, are factors to estimate,
$\U_1, \dots, \U_J \in \mathrm{St}_{d, k}$ are orthonormal subspace bases to estimate
($\mathrm{St}_{d, k}$ denotes the $d \times k$ Stiefel manifold),
$\z_i \sim \mathbb{C}\Nor(\0_k, \I_k)$ are \iid subspace coefficients,
and $\beps_i \sim \mathbb{C}\Nor(\0_d, \I_d)$ are \iid noise vectors.

The model
\eqref{eq:collas}
assumes all signals in subspace $j$ have the same texture value $\tau_j$.
That assumption is somewhat analogous
to how the MPPCA model assumes all samples in mixture $j$
have the same noise variance $v_j$.
Our proposed HeMPPCAT model instead allows samples in the same mixture component 
to have different noise variances,
and allows different signal components
to have different signal strengths,
rather than a common scaling factor
$\sqrt{\tau}$.
\comment{ 
, or "power levels." \textcolor{red}{AX: are differences mentioned after this point significant enough?} Additionally, all signals are assumed to lie in subspaces, meaning they must have zero mean and the factors to estimate are restricted to the Stiefel manifold. HeMPPCAT (and MPPCA) is more general, as it models samples that have non-zero mean, and there are no restrictions on the factor matrices to estimate.
}


\vsp{1}
\subsection{Covariance-Guided MPPCA}
\vsp{1}

Covariance-Guided MPPCA (C-MPPCA) \cite{han2015covariance} is an MPPCA variant that estimates the factor matrices and noise variances using a pooled sample covariance matrix, rather than a global sample covariance matrix, in each GEM iteration. 

C-MPPCA uses the original MPPCA model \eqref{eq:mppca_model} for each sample. As a result, C-MPPCA also assumes all samples in mixture component $j$ share a unique noise variance $v_j$.

\vsp{1}
\subsection{Mixtures of Robust PPCA}
\vsp{1}

Mixtures of Robust PPCA (MRPPCA) \cite{archambeau2008mixtures} is another MPPCA variant that is more robust to outlier samples. It also models each sample using \eqref{eq:mppca_model}, but assumes $\z_i \sim t_{\nu_j}(\bm 0, \I_d)$ and $\beps_i \sim t_{\nu_j}(\bm 0, v_j^{-1} \I_d)$, where $t_\nu(\bmu, \bm \Lambda)$ denotes the multivariate $t$-distribution with degrees of freedom $\nu$, mean $\bmu$, and scale matrix $\bm \Lambda$. The degrees of freedoms $\nu_j$ are additional parameters to estimate.

Again, like MPPCA and C-MPPCA, MRPPCA assumes all samples in mixture component $j$ have variance $v_j$.


\vsp{1}
\section{HeMPPCAT Data Model} \label{sec:model}
\vsp{1}

We assume
there are $n_1 + \dots + n_L = n$ data samples in $\Re^d$
from $L$ different noise groups
with model
\begin{equation} \label{eq:model}
    \y_{\ell, i} = \F_j \zl + \bmu_j + \beps_{\ell, i} 
\end{equation}
for all $i \in \{1, \dots, n_\ell\}$, $\ell \in \{1, \dots, L\}$,
and some $j = j_i \in \{1, \dots, J\}$.
$\F_1, \dots, \F_J \in \Re^{d \times k}$ are unknown factor matrices to estimate
(\emph{not} constrained to the Stiefel manifold,
so different signal components
can have different amplitudes),
$\zl \sim \Nor(\0_k, \I_k)$ are \iid coefficients,
and $\bmu_1, \dots, \bmu_J \in \Re^d$ are unknown mean vectors to estimate.
To model heteroscedastic noise, we assume
$\beps_{\ell, i} \sim \Nor(\0_d, v_\ell \I_d)$,
where $v_1, \dots, v_L$ are unknown noise variances to estimate. 
Importantly,
the noise model associates noise variance with data sample,
regardless of the underlying affine subspace.

The joint log-likelihood of the samples is
\begin{align} \label{eq:jll}
    \LL(\F, \bmu, \var, \p) &= \suml \sumi \lnrm \big\{p(\y_{\ell, i}) \big\}
    \nonumber\\
    &= \suml \sumi \lnrm \Bigg\{\sumj \pi_j p(\y_{\ell, i} \: | \: j) \Bigg\}, 
\end{align}
\vsp{1}
\begin{align*}
    p(\y_{\ell, i} \: | \: j) &= (2\pi)^{-d/2} \mathrm{det}(\C_{\ell, j})^{-1/2}\exprm(-\Frac{E_{\ell ij}^2}{2} ), \\
    E_{\ell ij}^2 &= (\y_{\ell, i} - \bmu_j)^T \C_{\ell, j}^{-1} (\y_{\ell, i} - \bmu_j), \\
    \C_{\ell, j} &= \F_j\FjT + v_{\ell}\I_d,
\end{align*}
\comment{
\begin{equation} \label{eq:jll}
\begin{gathered}
    \begin{aligned}
        \LL(\F, \bmu, \var, \p) &= \suml \sumi \lnrm \big\{p(\y_{\ell, i}) \big\} \\
        &= \suml \sumi \lnrm \Bigg\{\sumj \pi_j p(\y_{\ell, i} \: | \: j) \Bigg\}, 
    \end{aligned} \\
    p(\y_{\ell, i} \: | \: j) = (2\pi)^{-d/2} \mathrm{det}(\C_{\ell, j})^{-1/2}\exprm(-\Frac{E_{\ell ij}^2}{2} ), \\
    E_{\ell ij}^2 = (\y_{\ell, i} - \bmu_j)^T \C_{\ell, j}^{-1} (\y_{\ell, i} - \bmu_j), \\
    \C_{\ell, j} = \F_j\FjT + v_{\ell}\I_d,
\end{gathered}
\end{equation}
}
where $\F = [\F_1, \dots, \F_J]$, $\bmu = [\bmu_1, \dots, \bmu_J]$, $\var = [v_1, \dots, v_L]$,
and $\p = [\pi_1, \dots, \pi_J]$,
with $\pi_j$ being the $j$th \textit{mixing proportion},
such that $\pi_j \geq 0$ and $\sum_j \pi_j = 1$.
{
Conceptually,
to draw samples from this model,
one first picks a class $j$
according to the categorical distribution
with parameters $\p$,
and then picks random coefficients
to multiply the factor matrix $\F_j$,
and then adds noise,
where the noise variance
depends on the noise group.
}

\vsp{1}
\section{
GEM algorithm
} \label{sec:em}
\vsp{1}

The joint log-likelihood 
\eqref{eq:jll} is nonconvex with respect to 
the parameters of interest.
MPPCA \cite{tipping:99:mop}
maximized a similar expression
using an EM method.
We derived
a GEM
algorithm
to maximize \eqref{eq:jll} with respect to the parameters \F, \bmu, $\var$, and $\p$,
using a natural
\textit{complete-data log-likelihood} formulation. 

Let \zlj denote the coefficients associated with mixture $j$ for sample $\y_{\ell, i}$,
and let $g_{\ell ij}$
denote random variables having a
{categorical distribution}
where $g_{\ell ij} = 1$ indicates ``mixture $j$ generated sample $\y_{\ell, i}$.''
Treating $\zlj$ and $g_{\ell ij}$ as missing data,
the complete-data log-likelihood is
\begin{equation} \label{eq:cdll} 
\begin{gathered}
    \LL_C(\btheta) = \suml \sumi \sumj g_{\ell ij} \lnrm \big\{\pi_j p(\y_{\ell, i}, \zlj) \big\}, \\
    p(\y_{\ell, i}, \zlj) = p(\y_{\ell, i} | \zlj)p(\zlj)
    \\
    \y_{\ell, i} | \zlj
    \sim \Nor(\F_j\zlj + \bmu_j, v_\ell \I)
    \\
    p(\zlj) = (2\pi)^{-k/2}\exprm\{ -\frac{1}{2}\zlj^T \zlj \},
\end{gathered}
\end{equation}
where $\btheta = [\F, \bmu, \var, \p]$
is shorthand for all the parameters to estimate.

\vsp{1}
\subsection{Expectation step}
\label{ssec:exp}
\vsp{1}

For the E-step of the GEM algorithm,
we compute the expectation of \eqref{eq:cdll}
conditioned on the current iterate's parameter estimates.
Ignoring irrelevant constants,
one can show this expression is
\begin{align}
\label{eq:cdlle}
    \big\langle \LL_C(\btheta; \btheta^{(t)}) \big\rangle &=
    \suml \sumi \sumj R_{\ell ij}
    \Big\{ \lnrm(\pi_j) - \frac{d}{2}\lnrm(v_\ell)
    \nonumber\\
    & - \frac{1}{2} \trrm \Big( \zlC \Big) - \frac{1}{2v_\ell}\|\y_{\ell, i} - \bmu_j\|_2^2
    \nonumber\\
    & + \frac{1}{v_\ell} \zle^T \F_j^T (\y_{\ell, i} - \bmu_j)
    \nonumber\\
    & - \frac{1}{2v_\ell} \trrm \Big(\FjT \F_j \zlC \Big) \Big\},
\end{align}
where
$\btheta^{(t)} = [\Ft, \bmu^{(t)}, \vart, \p^{(t)}]$
denotes the parameter estimates at iteration $t$,
$\langle \cdot \rangle$ denotes conditional expectation,
the conditional moments $\zle$ and $\zlC$ are given by
\begin{equation} \label{eq:cm} 
    \begin{gathered}
        \zle = \Mljtinv \FjtT (\y_{\ell, i} - \bmu^{(t)}_j), \\
        \zlC = \varlt \Mljtinv + \zle \zle ^T, \\
        \Mljt = \varlt \I_k + \FjtT \Fjt,
    \end{gathered}
\end{equation}
and $R_{\ell ij}$ is the posterior \textit{mixing responsibility}
of mixture component $j$
generating sample $\y_{\ell, i}$:
\begin{equation} \label{eq:R} 
    R_{\ell ij}^{(t)} = \frac{p(\y_{\ell, i} \: | \: j) \, \pi_j}{p(\y_{\ell, i})},
\end{equation}
where the probabilities are evaluated
at the current parameter estimates
$\btheta^{(t)}$. 

\vsp{1}
\subsection{Maximization step}
\label{ssec:max}
\vsp{1}

For the M-step,
it appears impractical
to maximize \eqref{eq:cdlle}
over all parameters simultaneously,
so we adopt a
GEM 
approach
\cite{dempster:77:mlf}
where we update subsets of parameters
in sequence.

Maximizing \eqref{eq:cdlle}
with respect to $\pi_j$, $v_\ell$, $\bmu_j$, and $\F_j$,
in that sequence,
results in the following M-step update expressions
(derivations omitted due to page limits):
\begin{align}
    \pi^{(t+1)}_j &= \frac{1}{n} \sumliml \sumlimi \Rlijt
    \nonumber\\
    v^{(t+1)}_\ell &= \frac{1}{d \sumlimi \sumlimj \Rlijt} \Bigg[ \sumlimi \sumlimj \Rlijt \| \y_{\ell, i} - \bmu^{(t)}_j \|_2^2
    \nonumber\\ 
    &-2 \sumlimi \sumlimj \Rlijt \zle^T \FjtT (\y_{\ell, i} - \bmu^{(t)}_j)
    \nonumber\\
    &+ \sumlimi \sumlimj \Rlijt \trrm\bigg( \zlC \FjtT \Fjt \bigg) \Bigg]
    \nonumber\\
    \bmu^{(t+1)}_j &= \frac{\sumliml \sumlimi \frac{\Rlijt}{\varltnew}[\y_{\ell, i} - \Fjt \zle]}{\sumliml \sumlimi \frac{\Rlijt}{\varltnew}}
    \nonumber\\
    \F_j^{(t+1)} &= \Big( \sumliml \sumlimi \frac{\Rlijt}{v^{(t+1)}_{\ell}}
    (\y_{\ell, i} - \bmu^{(t+1)}_j) \zle^T \Big)
    \nonumber\\
    &\cdot \Big( \sumliml \sumlimi \frac{\Rlijt}{v^{(t+1)}_\ell} \zlC \Big)^{-1}
    .\nonumber
\end{align}
These expressions naturally generalize
those in \cite{tipping:99:mop}.
For the subsequent results,
we initialized the parameter estimates
by using final MPPCA estimates. MPPCA was initialized using 1000 iterations of $K$-Planes \cite{kambhatla1997dimension} \cite{bradley2000k}. 

\section{Experiments \& Results}
\label{sec:results}

\vsp{0.7}
\subsection{Synthetic Datasets}
\vsp{0.7}

We generated 25 separate synthetic datasets.
Each dataset contained $n = 10^3$ data samples of dimension $d = 10^2$
according to the model \eqref{eq:model},
where there were $L = 2$ noise groups and $k = 3$ factors
for each of $J = 3$ affine subspaces.
The factor matrices were generated as
$\F_j = \U_j \mathrm{Diag}^{1/2}(\bm \lambda)$ for $j = 1, \dots, J$,
where $\U_j \in \mathrm{St}_{d, k}$ was drawn uniformly at random,
and $\bm \lambda = (16, 9, 4)$ in all datasets.
The elements of the mean vectors $\bmu_j$ were drawn independently and uniformly at random in the interval $[0, 1]$. 

In all 25 datasets, the first $n_1 = 800$ samples had noise variance $v_1$
that we swept through from 1 to 4 in step sizes of 0.1.
Mixtures 1 and 2 each generated 250 of these samples,
while mixture 3 generated the remaining 300.
The other $n_2 = 200$ samples had noise variance $v_2 = 1$,
where mixtures 1 and 3 each generated 50 of these samples, and mixture 2 generated 100 of them.

We applied K-Planes,
MPPCA, and HeMPPCAT
to compute estimates
$\hat{\F}_j$
of the underlying factors
$\F_j$
across all 25 datasets.
In every dataset,
we recorded
the normalized estimation errors
\(
{\normF{ \hat{\F}_j\hat{\F}_j^T - \F_j\F_j^T }} / {\normF{ \F_j\F_j^T }}
\)
of all methods
at every value of $v_1$.
For each $v_1$,
we averaged the errors
across all 25 datasets. 

\begin{figure}[htb]
  \centering
  \centerline{\includegraphics[width=0.8\linewidth]{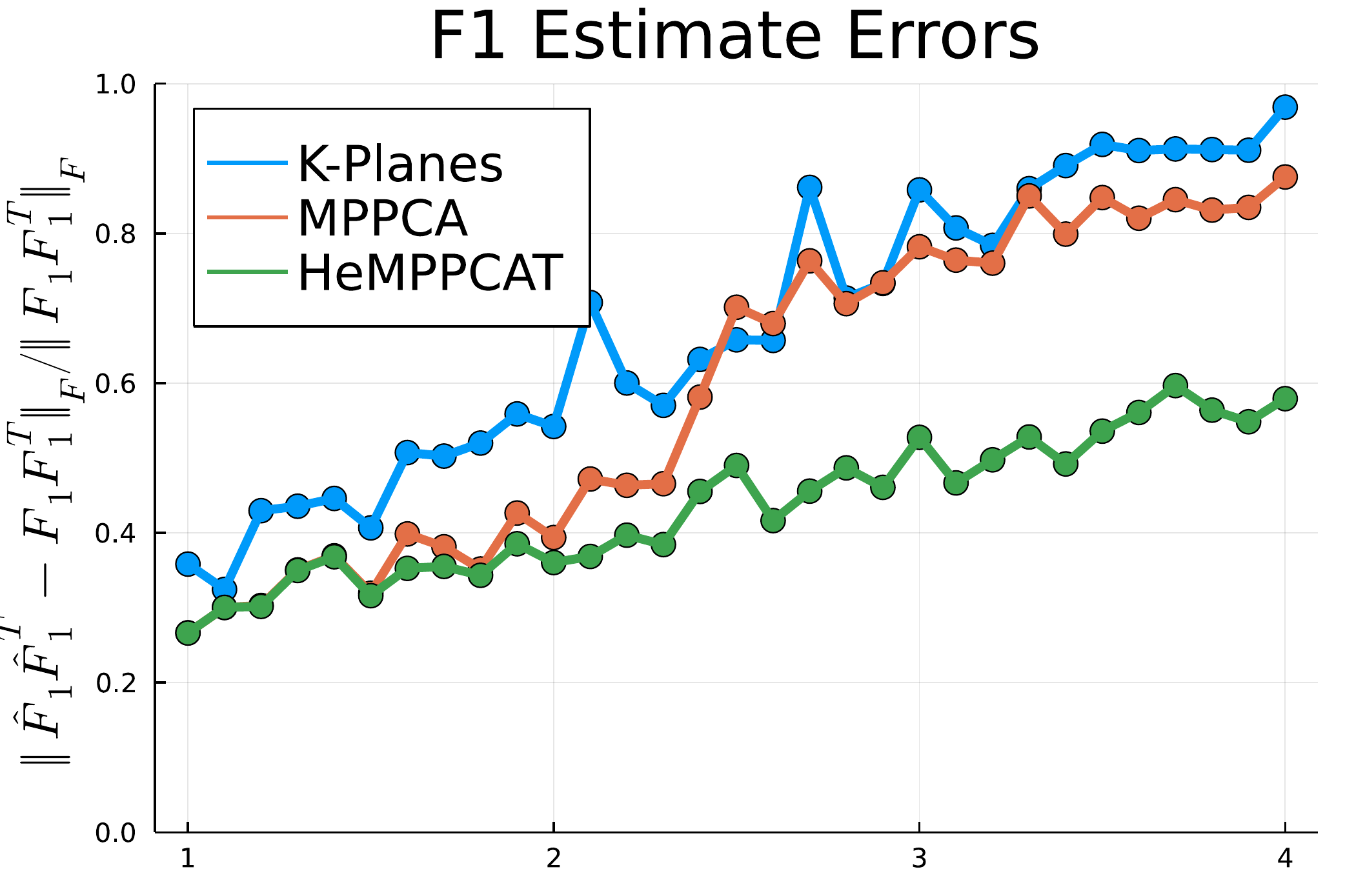}}
  (a) $\F_1$ error vs. $v_1$
\\[0.5em]
  \centerline{\includegraphics[width=0.8\linewidth]{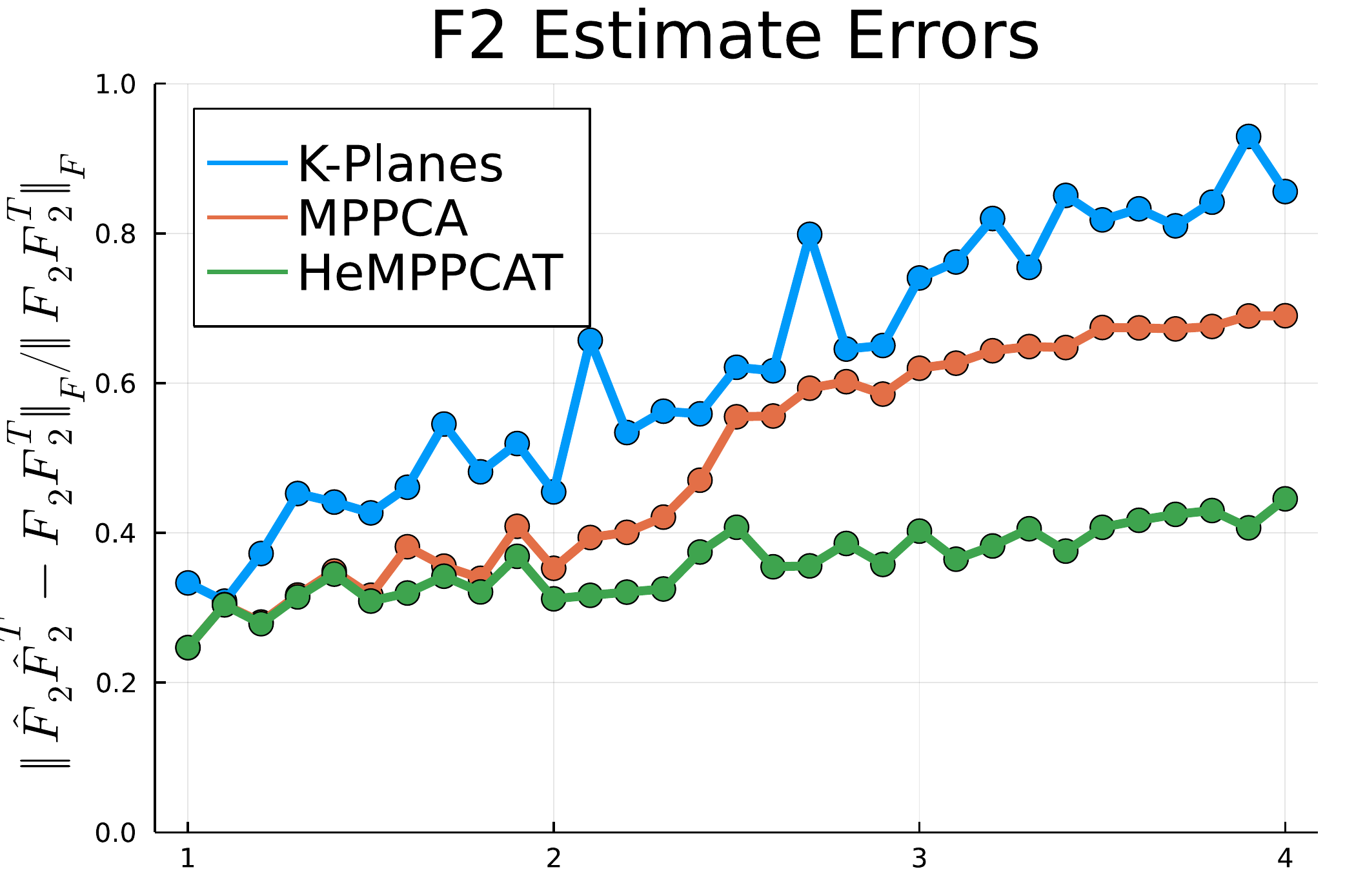}}
  (b) $\F_2$ error vs. $v_1$
  \\[0.5em]
  \centerline{\includegraphics[width=0.8\linewidth]{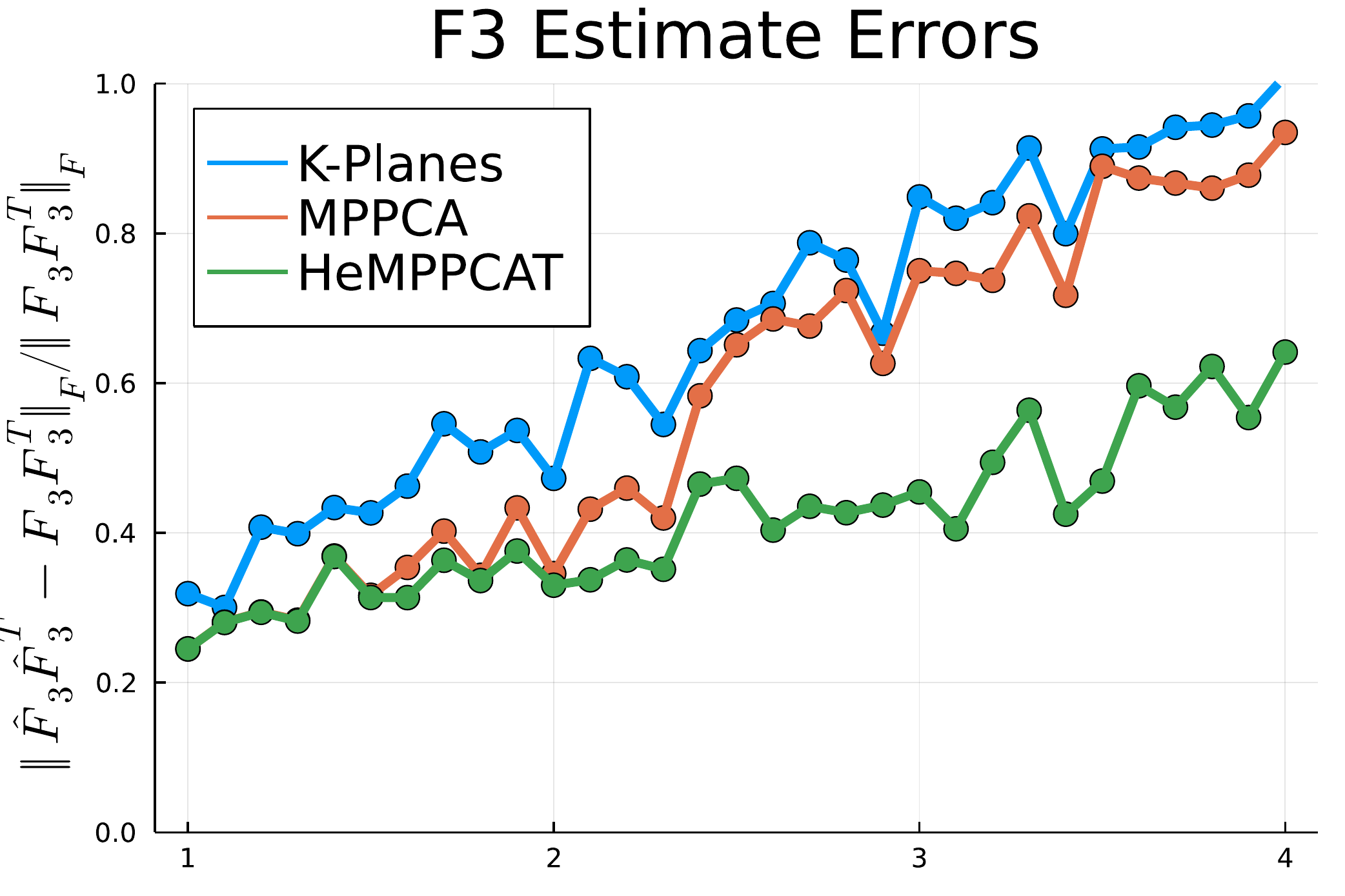}}
  \centerline{(c) $\F_3$ error vs. $v_1$}
\caption{Average $\F_j$ estimation error vs. $v_1$
(lower is better).}
\label{fig:F_synth}
\end{figure}

Figure~\ref{fig:F_synth} 
compares the average $\F_j$ estimation errors
across the 25 datasets against $v_1$.
When $v_1$ was close to $v_2$,
the dataset noise was fairly homoscedastic
and
MPPCA and HeMPPCAT had similar estimation errors.
As $v_1$ increased, the dataset noise became increasingly heteroscedastic,
and HeMPPCAT had much lower errors than  MPPCA. 
At almost all values of $v_1$,
the non-statistical K-Planes method
had higher errors than both MPPCA and HeMPPCAT.

\subsection{Hopkins 155 Dataset}
\vsp{0.7}

In computer vision, 
motion segmentation is 
the task of segmenting
moving objects in a 
video sequence into several 
independent regions, each 
corresponding to a different motion. 
The Hopkins 155 dataset is a 
series of 155 video sequences containing
bodies in motion. 
The dataset has the coordinates of 
$n$ feature points $\x_{f, i} \in \Re^2$
that were tracked across $F$ video frames.
A feature point \emph{trajectory} 
is formed by stacking 
the feature points across the frames: 
$\y_i = \begin{bmatrix} \x_{1, i}^T & \dots & \x_{F, i}^T \end{bmatrix}^T$. 
These trajectories are clustered
to segment the motions
using affine 
subspace clustering methods.
A single body's trajectories lie 
in an affine subspace with dimension 
of at most 4 \cite{tomasi1992shape},
so the trajectories of $J$ bodies 
lie in a union of
$J$ affine subspaces.
The dataset 
includes the ground-truth cluster
assignments for each trajectory.

Many motion segmentation methods
assume the moving bodies are \emph{rigid}.
In practice, many common objects of interest
do not satisfy 
rigid body motion.
For instance, feature points
on a walking human's legs 
move with different velocities
than feature points on the torso.
These differences can be modeled
as heteroscedastic noise: 
trajectories of the leg feature points 
may lie in a different noise group
than those of the torso. 


Each video sequence
contains either $J = 2$ or $J = 3$ moving bodies.
To simulate nonrigid body motion,
we added synthetic Gaussian noise 
to the trajectories.
In each sequence,
we created $L = 3$ synthetic 
noise groups with variances
$v_1, v_2, v_3$
corresponding to
signal to noise ratio (SNR) values of
$-30, -25, $ and $-20$ dB
relative to
that sequence's maximum trajectory 
$\ell_2$ norm.
Noise groups 1, 2, and 3 contained
50\%, 35\%, and 15\% 
of all trajectories, respectively.

For each sequence,
we divided the dataset of trajectories
into train and test sets
using an 80/20 split.
We applied K-Planes, MPPCA, and HeMPPCAT on the
train set, and then used 
parameter estimates
from all methods
to classify trajectories in the test set.
The test trajectories were
classified based on nearest affine subspace
using the K-Planes estimates,
and by
maximum likelihood
using MPPCA and HeMPPCAT estimates,
i.e., the 
predicted body for a test point $\y$ was 
$\argmax\limits_{j} \hat{\pi}_j p(\y \: | \: j ; \hat{\btheta})$, 
where $\hat{\pi}_j$ is the estimated mixing proportion 
of body $j$.
We computed
$p(\y \: | \: j ; \hat{\btheta})$
according to the approaches' 
respective data models 
\eqref{eq:mppca_model} and \eqref{eq:model}.


Table~\ref{table:hopkins} shows
the misclassification rate on 
the test set using the methods' parameter estimates.
Using HeMPPCAT's estimates 
achieved lower classification error
than using the other two approaches' estimates 
in each of the noise groups, and on
the overall test set.

\begin{table}
\adjustbox{width=\linewidth}{
\begin{tabular}{|c|c|c|c|}
    \hline
    & K-Planes & MPPCA & HeMPPCAT \\
    \hline
    Noise group 1 (low noise) & 24.1\% & 19.4\% & \textbf{18.6}\%\\
    \hline
    Noise group 2 (medium noise) & 24.5\% & 27.3\% & \textbf{19.3}\% \\
    \hline
    Noise group 3 (high noise) & 28.0\% & 34.8\% & \textbf{20.1}\% \\
    \hline
    Overall & 24.8\% & 24.5\% & \textbf{19.1}\% \\
    \hline
\end{tabular}
}
\caption{Average misclassification rate on Hopkins 155 video dataset
with synthetic heteroscedastic noise (lower is better).}
\label{table:hopkins}
\end{table}


\vsp{1.5}
\section{Conclusion}
\label{sec:concl}
\vsp{1.5}

This paper generalized MPPCA to jointly estimate
the underlying factors, means, and noise variances
from data with heteroscedastic noise.
The proposed EM algorithm sequentially updates the mixing proportions,
noise variances, means, and factor estimates.
Experimental results
on synthetic and the Hopkins 155 datasets
illustrate the benefit of accounting for heteroscedastic noise.

There are several possible interesting directions for future work.
We could generalize this approach even further
by accounting for other cases of heterogeneity,
e.g., missing data or heteroscedastic noise across features.
There may be faster convergence variants of EM
such as
a space-alternating generalized EM (SAGE) approach
\cite{fessler:94:sag}
that could be explored.
Another direction could be jointly estimating
the number of noise groups $L$ and mixtures $J$ along with the other parameters.

%% file: supp.tex

\clearpage
\onecolumn 

\appendix

\section{Complete-Data Log-Likelihood Expansion}
We first compute the expression for $p(\yli, \zlj) = p(\yli\: | \:\zlj) p(\zlj)$. Given the coefficients $\zlj$, the only random component in sample $\yli$ is due to the Gaussian noise vector $\beps_{\ell, i} \sim \Nor(\bm 0_k, \vl \I)$:
\begin{align}
    p(\yli\: | \:\zlj) &= (2\pi)^{-d/2} \detrm(\vl \I)^{-1/2} \exprm\Big\{  -\frac{1}{2} (\yli - \F_j \zlj - \bmu_j)^T (\vl \I)^{-1} (\yli - \F_J \zlj - \bmu_j) \Big\} \\
    &= (2\pi \vl)^{-d/2} \exprm \Big\{ -\frac{1}{2 \vl} \|\yli - \F_j - \zlj\|_2^2 \Big\}.
\end{align}
The coefficients $\zlj$ are assumed to be standard Gaussian vectors:
\begin{align}
    p(\zlj) &= (2\pi)^{-k/2} \detrm(\I)^{-1/2} \exprm \Big\{ -\frac{1}{2} \zlj^T \I^{-1} \zlj \Big\} \\
    &= (2\pi)^{-k/2} \exprm\Big\{-\frac{1}{2}\zlj^T\zlj\Big\}.
\end{align}
Multiplying the two expressions above results in the following joint distribution:
\begin{align} \label{eq:joint-prob}
    p(\yli, \zlj) = (2\pi \vl)^{-d/2} \exprm \Big\{ -\frac{1}{2 \vl} \|\yli - \F_j \zlj - \bmu_j \|_2^2 \Big\} (2\pi)^{-k/2} \exprm \Big\{ -\frac{1}{2} \zlj^T \zlj \Big\}.
\end{align}

Substituting \eqref{eq:joint-prob} into \eqref{eq:cdll} results in the following complete-data log-likelihood expression:
\begin{align}
    \nonumber
    \LL&_C(\theta) = \sumlij \glij \bigg\{ \lnrm\big\{ \pi_j p(\yli, \zlj) \big\} \bigg\} \\
    \nonumber
    &= \sumlij \glij \bigg\{ \lnrm (\pi_j) + \lnrm \big( p(\yli, \zlj) \big) \bigg\} \\
    \nonumber
    &= \sumlij \glij \bigg\{ \lnrm (\pi_j)  - \frac{d}{2} \lnrm(\vl) - \frac{1}{2\vl} \|\yli - \F_j \zlj - \bmu_j\|_2^2 - \frac{1}{2}\zlj^T\zlj \bigg\} \\
    \nonumber
    &= \sumlij \glij \bigg\{ \lnrm(\pi_j) - \frac{d}{2}\lnrm(\vl) - \frac{1}{2\vl}(\yli - \F_j \zlj - \bmu_j)^T(\yli - \F_j \zlj - \bmu_j) - \frac{1}{2}\zlj^T\zlj \bigg\} \\
    \nonumber
    &= \sumlij \glij \bigg\{ \lnrm(\pi_j) - \frac{d}{2}\lnrm(\vl) - \frac{1}{2}\zlj^T\zlj - \frac{1}{2\vl}\|\yli - \bmu_j\|_2^2 + \frac{1}{\vl} \zlj^T \FjT (\yli - \bmu_j) - \frac{1}{2\vl} \zlj^T\FjT\F_j\zlj  \bigg\} \\
    \nonumber
    &= \sumlij \glij \bigg\{ \lnrm(\pi_j) - \frac{d}{2}\lnrm(\vl) - \frac{1}{2}\trrm(\zlj\zlj^T) - \frac{1}{2\vl}\|\yli - \bmu_j\|_2^2 + \frac{1}{\vl} \zlj^T \FjT (\yli - \bmu_j) - \frac{1}{2\vl} \trrm(\FjT\F_j\zlj\zlj^T)  \bigg\}.
\end{align}
We ignored the constant terms $-\frac{d}{2} \lnrm(2\pi)$ and $-\frac{k}{2} \lnrm(2\pi)$.

\newpage

\section{Expectation Step Derivation}
\subsection{Product of Conditional Expectations}
The E-Step of EM requires computing the expectation of \eqref{eq:cdll} conditioned on the data samples $\yli$ for $i = 1, \dots, n_\ell$ and $\ell = 1, \dots, L$. The expected value of $\LL_C(\theta)$ given the samples is
\begin{align} \label{eq:cdlle-expanded}
    \langle \LL_C(\theta) \rangle = \sumlij \: &\expec\Bigg[ \glij \lnrm(\pi_j) - \glij \frac{d}{2}\lnrm(\vl) - \glij \frac{1}{2}\trrm(\zlj\zlj^T) \\
    \nonumber
    &- \glij \frac{1}{2\vl}\|\yli - \bmu_j\|_2^2 + \glij \frac{1}{\vl} \zlj^T \FjT (\yli - \bmu_j) - \glij \frac{1}{2\vl} \trrm(\FjT\F_j\zlj\zlj^T) \: \bigg| \: \yli \Bigg]. 
\end{align}

To simplify \eqref{eq:cdlle-expanded}, we show $\expec\big[ \glij \zlj\: | \:\yli \big] = \expec\big[ \glij\: | \:\yli \big] \expec\big[ \zlj\: | \:\glij = 1, \yli \big]$ for all $i = 1, \dots, n_\ell$, $\ell = 1, \dots, L$, and $j = 1, \dots, J$:
\begin{align}
    \nonumber
    \expec\Big[ &\glij \zlj\: | \:\yli \Big] = \expec \Big[ \glij \expec\big[ \zlj\: | \:\glij, \yli \big] \big| \yli \Big] \\
    \nonumber
    &= \bigg( 0 * \expec\Big[\zlj\: | \:\glij = 0, \yli \Big]*p(\glij = 0\: | \:\yli) \bigg) + \bigg( 1*\expec\Big[\zlj\: | \:\glij = 1, \yli \Big] * p(\glij = 1\: | \:\yli) \bigg) \\ 
    \nonumber
    &= \expec\Big[\zlj\: | \:\glij = 1, \yli \Big] \expec\Big[ \glij\: | \:\yli \Big].
\end{align}

Therefore:
\begin{align}
    \nonumber
    \langle \LL_C(\theta) \rangle = \sumlij \: &\expec\Big[ \glij\: | \:\yli \Big]\lnrm(\pi_j) - \frac{d}{2} \expec\Big[ \glij\: | \:\yli \Big] \lnrm(\vl) -  \frac{1}{2} \expec\Big[ \glij\: | \:\yli \Big] \trrm\Big(\expec\big[ \zlj \zlj^T\: | \:\glij = 1, \yli \big] \Big) \\ 
    \nonumber
    &- \frac{1}{2\vl} \expec\Big[ \glij\: | \:\yli \Big]  \|\yli - \bmu_j\|_2^2 + \frac{1}{\vl} \expec\Big[ \glij\: | \:\yli \Big] \expec\Big[ \zlj\: | \:\glij = 1, \yli \Big]^T \FjT(\yli - \bmu_j) \\
    \nonumber
    &- \frac{1}{2\vl} \expec\Big[ \glij\: | \:\yli \Big]
    \trrm\Big( \FjT \F_j \expec\big[ \zlj \zlj^T\: | \:\glij = 1, \yli \big] \Big).
\end{align}

Letting $R_{\ell ij} := \expec[\glij\: | \:\yli]$, $\zle := \expec[\zlj\: | \:\glij = 1, \yli]$, and $\zlC := \expec[\zlj \zlj^T\: | \:\glij = 1, \yli]$, the expected complete-data log-likelihood is
\begin{align} 
    \nonumber
    \langle \LL_C(\theta) \rangle = \sumlij &R_{\ell ij} \bigg\{ \lnrm(\pi_j) - \frac{d}{2} \lnrm(\vl) -  \frac{1}{2} \trrm\Big( \zlC \Big) \\ 
    \nonumber
    &- \frac{1}{2\vl} \|\yli - \bmu_j\|_2^2 + \frac{1}{\vl} \zle^T \FjT(\yli - \bmu_j) \\
    \nonumber
    &- \frac{1}{2\vl}
    \trrm\Big( \FjT \F_j \zlC \Big) \bigg\}.
\end{align}

\subsection{Mixing Responsibilities}
Since $\glij$ is a categorical random variable, the expectation $R_{\ell ij}$ is simply $p(\glij = 1\: | \:\yli)$. This expression can be computed using Bayes' rule:
\begin{align}
    R_{\ell ij} = \expec[\glij\: | \:\yli] = p(\glij = 1\: | \:\yli) = \frac{p(\yli\: | \:\glij = 1) p(\glij = 1)}{p(\yli)} = \frac{p(\yli\: | \:j) \pi_j}{p(\yli)}.
\end{align}

\subsection{Conditional Coefficent Expectation and Covariance}
To derive $\zle$ and $\zlC$, we first compute the conditional probability distribution $p(\zlj\: | \:\glij = 1, \yli)$. This can again be done using Bayes' rule:
\begin{align}
    \nonumber
    p&(\zlj\: | \:\glij = 1, \yli) = \frac{p(\yli\: | \:\glij = 1, \zlj) p(\zlj\: | \:\glij = 1)}{p(\yli\: | \:\glij = 1)} \\
    \nonumber
    &= \frac{ (2\pi\vl)^{-d/2} \exprm \Big\{ -\frac{1}{2\vl} \|\yli - \F_j \zlj - \bmu_j\|_2^2 \Big\} (2\pi)^{-k/2} \exprm\Big\{ -\frac{1}{2} \|\zlj\|_2^2 \Big\} }{ (2\pi)^{-d/2} \detrm(\C_j)^{-1/2} \exprm\Big\{ -\frac{1}{2} (\yli - \bmu_j)^T \Cljinv (\yli - \bmu_j) \Big\} } \\
    \nonumber
    &= (2\pi)^{-k/2} \vl^{-d/2} \detrm(\Clj)^{1/2} \exprm\Big\{ -\frac{1}{2\vl} \|\yli - \F_j\zlj - \bmu_j\|_2^2 - \frac{1}{2}\|\zlj\|_2^2 + \frac{1}{2} (\yli - \bmu_j)^T \Cljinv (\yli - \bmu_j) \Big\}.
\end{align}
The term $\vl^{-d/2} \detrm(\Clj)^{1/2}$ can be simplified as such:
\begin{align}
    \nonumber
    \vl^{-d/2} &\detrm(\Clj)^{1/2} = \Big( \vl^{-d} \detrm(\Clj) \Big)^{1/2} \\
    \nonumber 
    &= \Big( \vl^{-d} \detrm(\F_j \FjT + \vl \I_d) \Big)^{1/2} \\
    \nonumber
    &= \Big( \vl^{-d} \detrm\big( \I_k + \FjT(\vl \I_d)^{-1} \F_j \big) \detrm(\vl \I_d)  \Big)^{1/2} \\
    \nonumber
    &= \Big( \vl^{-d} \detrm\big( \I_k + \vl^{-1} \FjT \F_j \big) \vl^d \Big)^{1/2} \\
    \nonumber
    &= \detrm\big( \vl^{-1} (\vl\I_k + \FjT \F_j) \big)^{1/2} \\
    \nonumber
    &= \detrm\big( \vl^{-1} \Mlj \big)^{1/2} = \detrm\big( \vl \Mljinv \big)^{-1/2}.
\end{align}

The term inside the $\exprm(\cdot)$ can also be simplified as such. Let $\yblij := (\yli - \bmu_j)$:
\begin{align}
    \nonumber
    -\frac{1}{2\vl} &\|\yblij - \F_j\zlj\|_2^2 - \frac{1}{2} \|\zlj\|_2^2 + \frac{1}{2}\yblij^T\Cljinv\yblij \\
    \nonumber
    &= -\frac{1}{2\vl} (\yblij - \F_j\zlj)^T(\yblij - \F_j\zlj) - \frac{1}{2} \zlj^T\zlj + \frac{1}{2}\yblij^T\Cljinv \yblij \\
    \nonumber
    &= -\frac{1}{2\vl} \bigg[ \yblij^T\yblij - \yblij^T \F_j\zlj - \zlj^T\FjT \yblij + \zlj^T\FjT\F_j\zlj + \vl\zlj^T\zlj - \vl \yblij^T \Cljinv \yblij \bigg] \\
    \nonumber
    &= -\frac{1}{2\vl} \bigg[ \yblij^T\yblij - \yblij^T \F_j\zlj - \zlj^T\FjT \yblij + \zlj^T\FjT\F_j\zlj + \vl\zlj^T\zlj - \yblij^T (\I_d - \F_j \Mljinv \FjT) \yblij \bigg] \\
    \nonumber
    &= -\frac{1}{2\vl} \bigg[ \yblij^T (\F_j\Mljinv\FjT) \yblij - \yblij^T \F_j\zlj - \zlj^T\FjT \yblij + \zlj^T\Mlj\zlj \bigg] \\
    \nonumber
    &= -\frac{1}{2\vl} \bigg[ \zlj^T\Mlj\zlj - \zlj^T\FjT \yblij - \yblij^T \F_j \Mljinv \Mlj \zlj + \yblij^T (\F_j\Mljinv\FjT) \yblij \bigg] \\
    \nonumber
    &= -\frac{1}{2\vl} \bigg[ (\zlj - \Mljinv \F_j \yblij)^T \Mlj (\zlj - \Mljinv \F_j \yblij) \bigg] \\
    \nonumber
    &= -\frac{1}{2} \bigg[ (\zlj - \Mljinv \F_j \yblij)^T (\vl \Mljinv)^{-1} (\zlj - \Mljinv \F_j \yblij) \bigg],
\end{align}
where $\Cljinv = \vl^{-1}(\I_d - \F_j \Mljinv \FjT)$ is derived using the Matrix-Inversion Lemma. \\

The conditional probability distribution $p(\zlj\: | \:\glij=1, \yli)$ is equal to
\begin{align} \label{eq:cond-pdf}
    (2\pi)^{-k/2} \detrm(\vl \Mljinv)^{-1/2} \exprm\Big( -\frac{1}{2} (\zlj - \Mljinv \F_j \yblij)^T (\vl \Mljinv)^{-1} (\zlj - \Mljinv \F_j \yblij) \Big)
\end{align}
which is a Gaussian distribution with mean $\Mljinv \F_j (\yli - \bmu_j)$ and covariance $\vl \Mljinv$.

The conditional expectation $\zle$ is therefore $\Mljinv \F_j (\yli - \bmu_j)$. The term $\zlC$ can be derived as follows:
\begin{align}
    \nonumber
    \vl\Mljinv &= \expec[(\zlj - \zle)(\zlj - \zle)^T \: | \: \glij = 1, \yli] \\ 
    \nonumber
    &= \expec[\zlj\zlj^T \: | \: \glij = 1, \yli] - \expec[\zlj \: | \: \glij = 1, \yli] \: \expec[\zlj \: | \: \glij = 1, \yli]^T \\
    \nonumber
    &= \zlC - \zle \zle^T \\
    \zlC &= \vl \Mljinv + \zle \zle^T.
\end{align}

\subsection{Final E-Step Computation}
The final expected complete data log-likelihood is 
\begin{align} \label{eq:cdlle-proof}
    \langle \LL_C(\theta; \theta^{(t)}) \rangle = \sumlij &R_{\ell ij}^{(t)} \bigg\{ \lnrm(\pi_j) - \frac{d}{2} \lnrm(\vl) -  \frac{1}{2} \trrm\Big( \zlC \Big) \\ 
    \nonumber
    &- \frac{1}{2\vl} \|\yli - \bmu_j\|_2^2 + \frac{1}{\vl} \zle^T \FjT(\yli - \bmu_j) \\
    \nonumber
    &- \frac{1}{2\vl}
    \trrm\Big( \FjT \F_j \zlC \Big) \bigg\},
\end{align}
where the conditional expectations are computed using the current EM iterate's parameter estimates:
\begin{equation} \label{eq:cm-proof}
\begin{gathered}
    \zle = \Mljtinv \FjtT(\yli - \bmu_j^{(t)}) \\
    \zlC = \varlt \Mljtinv + \zle \zle^T \\
    \M_{\ell, j}^{(t)} = \varlt \I_k + \FjtT \Fjt.
\end{gathered}
\end{equation}

\newpage

\section{Maximization Step Derivation}
\subsection{Mixing Proportions}
We derive the update expression for each mixing proportion $\pi_j$ by maximizing \eqref{eq:cdlle-proof} with respect to $\pi_j$ subject to the constraint $\sumlimj \pi_j = 1$. This is equivalent to maximizing
\begin{equation} \label{eq:pi-opt}
    \sumlimj \sumliml \sumlimi R_{\ell ij}^{(t)} \lnrm(\pi_j)
\end{equation}
subject to the same constraint. To account for this constraint, we introduce the Lagrange multiplier $\lambda$. Let $f_1 := \sumlimj R_j \lnrm(\pi_j) + \lambda \Big( \sumlimj \pi_j - 1 \Big)$. Suppose we choose an arbitrary $j \in \{1, \dots, J\}$. We first calculate an expression for $\pitnew$ in terms of $\lambda$:
\begin{equation} \nonumber
\begin{gathered}
    \frac{\partial f_1}{\partial \pi_j} = \frac{\sumliml \sumlimi R_{\ell ij}^{(t)}}{\pi_j} + \lambda, \\
    \frac{\sumliml \sumlimi R_{\ell ij}^{(t)}}{\pitnew} + \lambda = 0, \\
    \pitnew = -\frac{\sumliml \sumlimi R_{\ell ij}^{(t)}}{\lambda}. 
\end{gathered}
\end{equation}

We then use the constraint function to find an expression for $\lambda$ in terms of $R_{\ell ij}^{(t)}$, and use that to calculate the final expression for $\pitnew$:
\begin{equation}
\nonumber
\begin{gathered}
    \sumlimj \pitnew = -\sumlimj \frac{\sumliml \sumlimi R_{\ell ij}^{(t)}}{\lambda} = 1, \\
    \lambda = -\sumlimj \sumliml \sumlimi R_{\ell ij}^{(t)} = - n, \\
    \pitnew = \frac{\sumliml \sumlimi R_{\ell ij}^{(t)}}{n}.
\end{gathered}
\end{equation}
This matches the update expression in Section \ref{ssec:max} and holds for all $j \in \{1, \dots, J\}$.

\subsection{Noise Variance}
Deriving the update expression for each variance $\vl$ is equivalent to maximizing
\begin{equation}\label{eq:var-opt}
    f_2 = \sumlij R_{\ell ij} \bigg\{ -\frac{d}{2}\lnrm(\vl) - \frac{1}{2\vl}\|\yli - \bmu\|_2^2 + \frac{1}{\vl} \zle^T\FjT (\yli - \bmu) - \frac{1}{2\vl}\trrm(\FjT\F_j\zlC) \bigg\}
\end{equation}
with respect to $\vl$. Choosing an arbitrary $\ell \in \{1, \dots, L\}$:
\begin{equation}
\nonumber
\begin{gathered}
    \frac{\partial f_2}{\partial \vl} = \sumlimi \sumlimj R_{\ell ij}^{(t)} \bigg\{ -\frac{d}{2\vl} + \frac{1}{2\vl^2}\|\yli\|_2^2 - \frac{1}{\vl^2} \zle^T \FjT (\yli - \bmu_j) + \frac{1}{2\vl^2} \trrm(\FjT\F_j \zlC) \bigg\}, \\
    \frac{1}{\varltnew} \sumlimi \sumlimj R_{\ell ij} \bigg\{ -\frac{d}{2} + \frac{1}{2\varltnew}\|\yli - \bmu_j\|_2^2 - \frac{1}{\varltnew} \zle^T \FjT (\yli - \bmu_j) + \frac{1}{2\varltnew} \trrm(\FjT\F_j \zlC) \bigg\} = 0, \\
    \sumlimi \sumlimj R_{\ell ij}^{(t)} \bigg\{ -\frac{d}{2} + \frac{1}{2\varltnew}\|\yli - \bmu_j\|_2^2 - \frac{1}{\varltnew} \zle^T \FjT (\yli - \bmu_j) + \frac{1}{2\varltnew} \trrm(\FjT\F_j \zlC) \bigg\} = 0, \\
    \frac{1}{\varltnew} \sumlimi \sumlimj R_{\ell ij} \bigg\{ \frac{1}{2}\|\yli - \bmu_j\|_2^2 - \zle^T \FjT (\yli - \bmu_j) + \frac{1}{2} \trrm(\FjT \F_j \zlC) \bigg\} = \frac{d}{2} \sumlimi \sumlimj R_{\ell ij}^{(t)}, \\
    \varltnew = \frac{1}{d \sumlimi \sumlimj R_{\ell ij}^{(t)}} \Bigg[ \sumlimi \sumlimj R_{\ell ij} \bigg\{ \|\yli - \bmu_j\|_2^2 - 2 \zle^T \FjT (\yli - \bmu_j) + \trrm(\FjT \F_j \zlC) \bigg\} \Bigg].
\end{gathered}
\end{equation}
This holds for all $\ell \in \{1, \dots, L\}$. Using the current iterate's estimates $\bmu_j^{(t)}$ and $\Fjt$ for $\bmu_j$ and $\F_j$, respectively, makes the expression match the corresponding equation in Section \ref{ssec:max}.

\subsection{Means}
Deriving the update expression for each mean $\bmu_j$ is equivalent to maximizing
\begin{equation}\label{eq:mu-opt}
    f_3 = \sumlimj \sumliml \sumlimi R_{\ell ij} \bigg\{-\frac{1}{2\vl}\|\yli - \bmu\|_2^2 + \frac{1}{\vl} \zle^T\FjT (\yli - \bmu) \bigg\}
\end{equation}
with respect to $\bmu_j$. Choosing an arbitrary $j \in \{1, \dots, J\}$:
\begin{equation}
\nonumber
\begin{gathered}
    \nabla_{\bmu_j} f_3 = \sumliml \sumlimi \frac{R_{\ell ij}}{\vl} \bigg\{ \yli - \bmu_j - \F_j \zlj \bigg\}, \\
    \sumliml \sumlimi \frac{R_{\ell ij}}{\vl} \bigg\{ \yli - \bmujtnew - \F_j \zlj \bigg\} = 0, \\
    \bmujtnew = \frac{\sumliml \sumlimi \frac{R_{\ell ij}^{(t)}}{\vl}[\yli - \F_j \zle]}{\sumliml \sumlimi \frac{R_{\ell ij}^{(t)}}{\vl}}.
\end{gathered}
\end{equation}
This holds for all $j \in \{1, \dots, J\}$. Using the current iterate's estimate $\F_j^{(t)}$ for $\F_j$, and the new estimate $\varltnew$ for $\vl$, makes the expression match the corresponding equation in Section \ref{ssec:max}.

\subsection{Factor Matrices}
Deriving the update expression for each factor matrix $\F_j$ is equivalent to maximizing
\begin{equation} \label{eq:F-opt}
    f_4 = \sumlimj \sumliml \sumlimi R_{\ell ij} \bigg\{\frac{1}{\vl} \zle^T\FjT (\yli - \bmu) - \frac{1}{2\vl} \trrm(\FjT\F_j \zlC) \bigg\}
\end{equation}
with respect to $\F_j$. Choosing an arbitrary $j \in \{1, \dots, J\}$:
\begin{align} 
    \nonumber
    f_4 &= \sumliml \sumlimi R_{\ell ij}^{(t)} \bigg\{\frac{1}{\vl} \zle^T\FjT (\yli - \bmu) - \frac{1}{2\vl} \trrm(\FjT\F_j \zlC) \bigg\} \\
    \nonumber
    &= \sumliml \sumlimi \frac{R_{\ell ij}^{(t)}}{\vl} \bigg\{ \trrm\Big( (\yli - \bmu_j)^T\F_j\zle \Big) - \trrm(\FjT \F_j \zlC) \bigg\} \\
    \nonumber
    &= \sumliml \sumlimi \frac{R_{\ell ij}^{(t)}}{\vl} \bigg\{ \trrm\Big( \F_j\zle (\yli - \bmu_j)^T \Big) - \trrm( \F_j \zlC \FjT) \bigg\} \\
    \nonumber
    &= \trrm\Bigg( \sumliml \sumlimi \frac{R_{\ell ij}^{(t)}}{\vl} \F_j \zle (\yli - \bmu_j)^T - \sumliml \sumlimi \frac{R_{\ell ij}^{(t)}}{\vl} \F_j \zlC \FjT \Bigg)\\
    \nonumber
    &= \trrm \Bigg( \F_j \sumliml \sumlimi \frac{R_{\ell ij}^{(t)}}{\vl}\zle (\yli - \bmu_j)^T - \frac{1}{2}\F_j \sumliml \sumlimi \frac{R_{\ell ij}^{(t)}}{\vl} \zlC \FjT \Bigg).
\end{align}
Let $\B_j := \sumliml \sumlimi \frac{R_{\ell ij}^{(t)}}{\vl} \zle (\yli - \bmu_j)^T$ and $\K_j := \sumliml \sumlimi \frac{R_{\ell ij}^{(t)}}{\vl} \zlC$.
\begin{align}
    \nonumber
    f_4 &= \trrm\big( \F_j \B_j - \frac{1}{2} \F_j \K_j \FjT \big)\\
    \nonumber
    &= \trrm\big( \F_j \K_j^{1/2} \K_j^{-1/2} \B_j - \frac{1}{2} \F_j \K_j^{1/2} \K_j^{1/2} \FjT \big),
\end{align}
where $\A^{1/2}$ denotes the principal square root of a positive semidefinite matrix $\A$, and $\A^{-1/2}$ denotes the inverse $\A^{1/2}$ if it is positive definite. Now let $\X_j = \F_j \K_j^{1/2}$ and $\tilde{\B}_j = \K_j^{-1/2} \B_j$.
\begin{equation}
\nonumber
\begin{gathered}
    f_4 = \trrm\big( \X_j \tilde{\B}_j - \frac{1}{2} \X_j\X_j^T \big), \\
    \nabla_{\X_j} f_4 = \tilde{\B}_j - \X_j, \\
    \tilde{\B}_j - \hat{\X}_j = 0, \\
    \Fjtnew \K_j^{1/2} = \B_j^T \K_j^{-1/2}, \\
    \Fjtnew = \B_j^T \K_j^{-1} = \bigg[ \sumliml \sumlimi \frac{R_{\ell ij}^{(t)}}{\vl} (\yli - \bmu_j) \zle^T \bigg] \bigg[ \sumliml \sumlimi \frac{R_{\ell ij}^{(t)}}{\vl} \zlC \bigg]^{-1}.
\end{gathered}
\end{equation}
This holds for all $j \in \{1, \dots, J\}$. Using the new estimates $\varltnew$ and $\bmujtnew$ for $\vl$ and $\bmu_j$, respectively, makes this expression match the corresponding equation in Section \ref{ssec:max}.

\newpage
\section{Additional Simulation Results}
\renewcommand{\thetable}{D\arabic{table}}
\renewcommand{\thefigure}{D\arabic{figure}}

\begin{figure}[htb]  
\minipage{0.32\textwidth}
    \includegraphics[width=\linewidth]{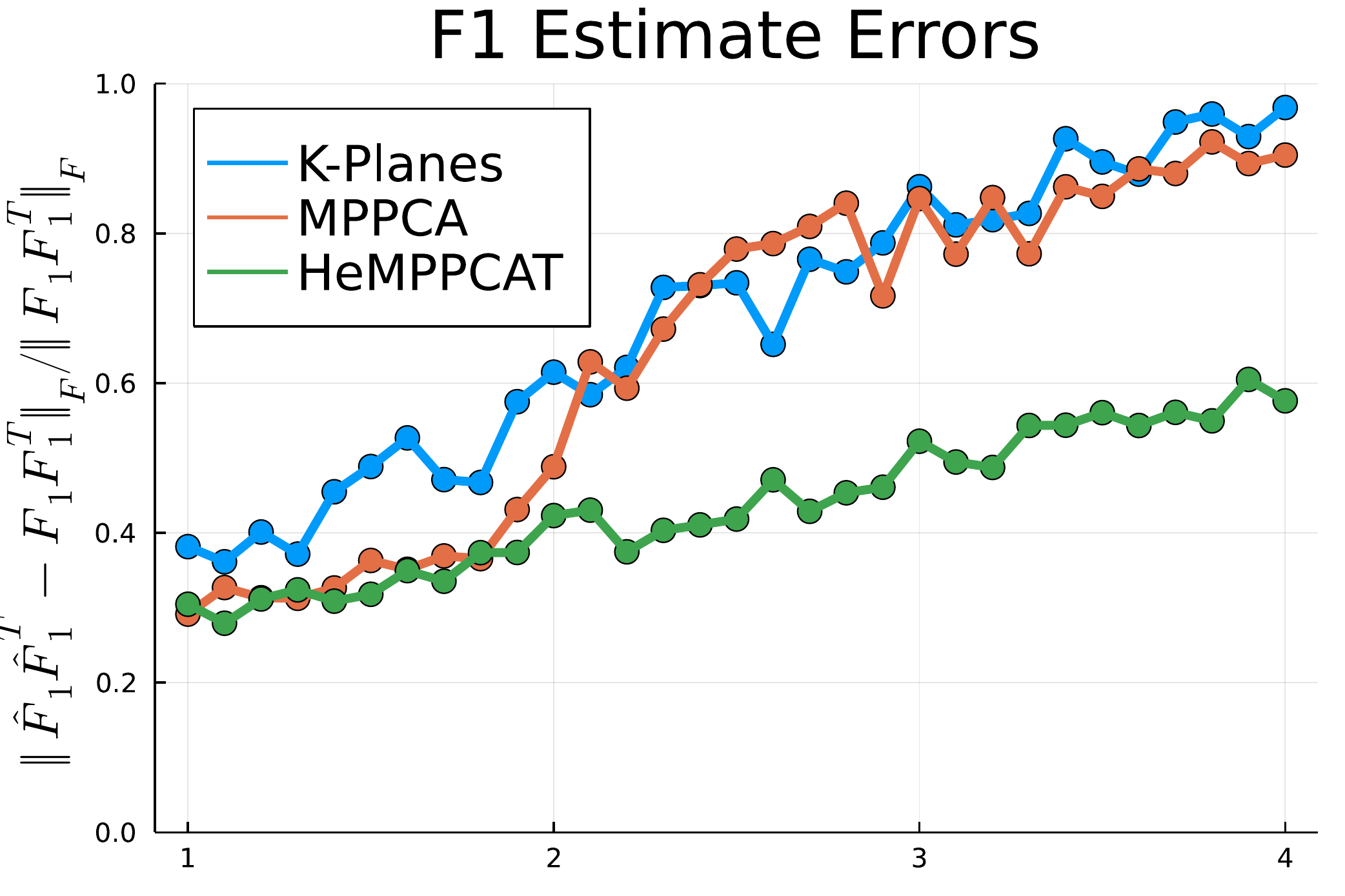}
    \centerline{(a) $\blmath{F}_1$ error vs. $v_1$}
\endminipage\hfill
\minipage{0.32\textwidth}
  \includegraphics[width=\linewidth]{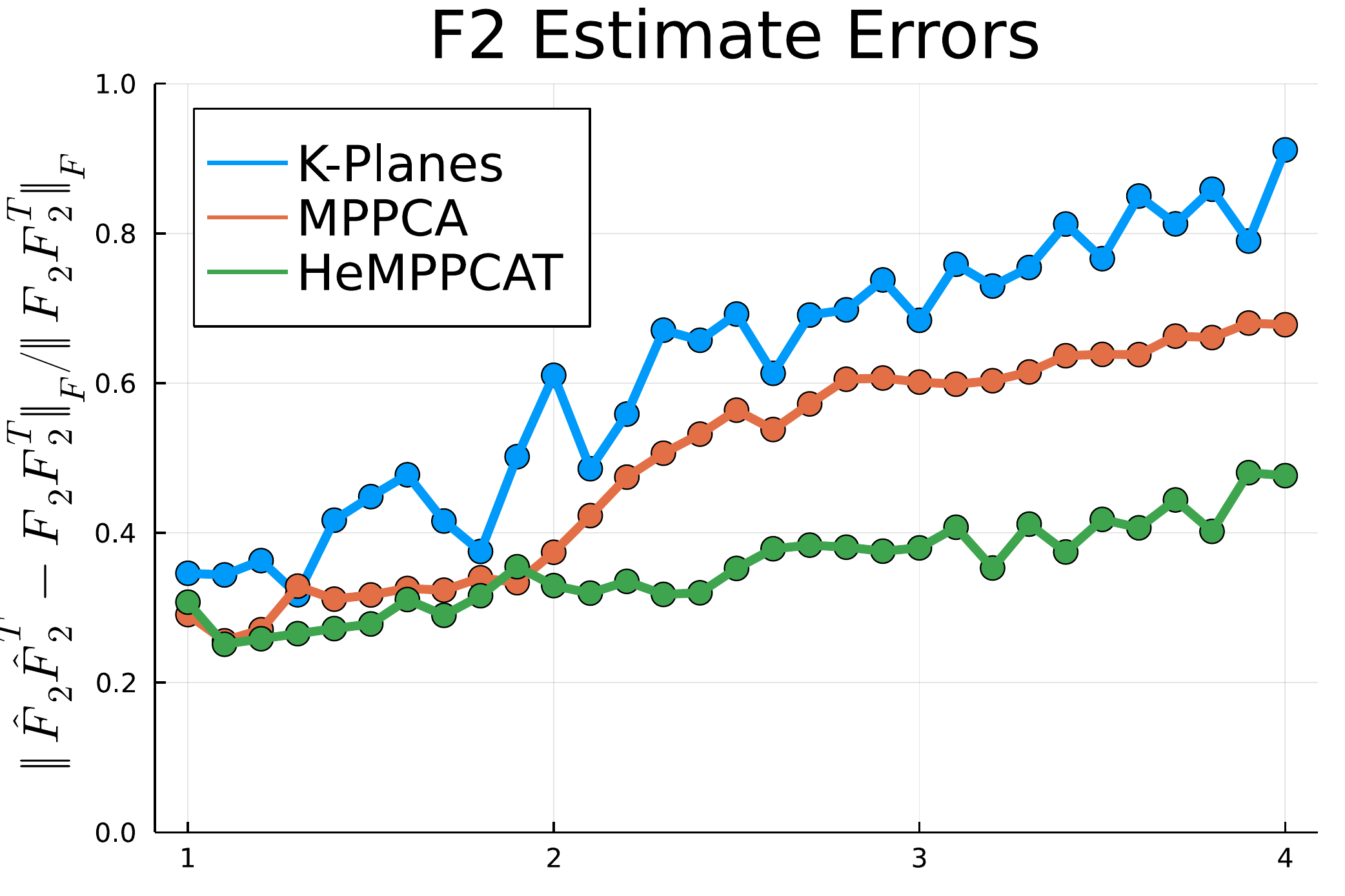}
  \centerline{(b) $\F_2$ error vs. $v_1$}
\endminipage\hfill
\minipage{0.32\textwidth}
  \includegraphics[width=\linewidth]{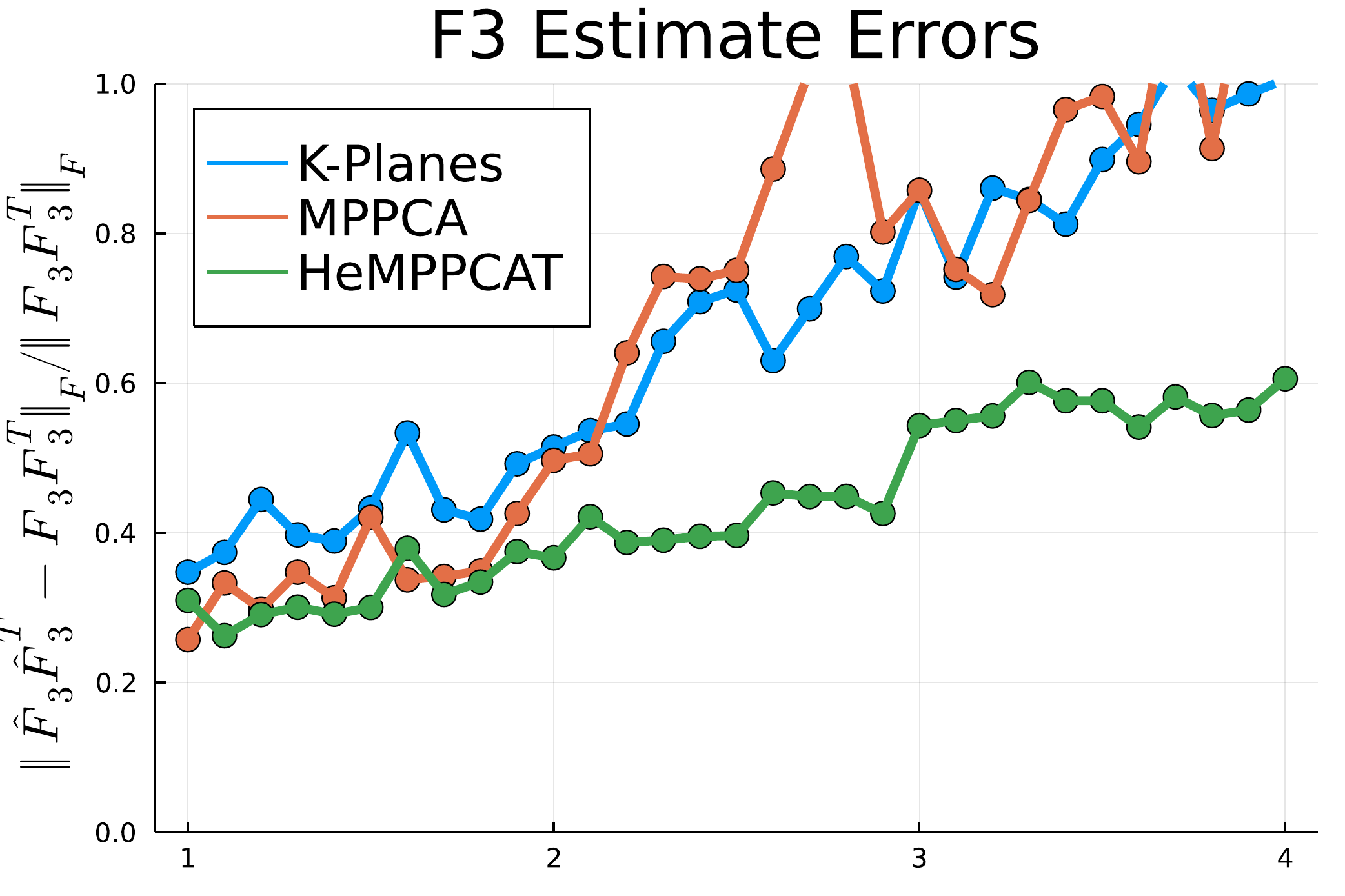}
  \centerline{(c) $\F_3$ error vs. $v_1$}
\endminipage\hfill

\caption{Average $\F_j$ estimation error vs. variance $v_1$.}
\label{fig:F_synth_supp}
\end{figure}

We repeated the synthetic dataset experiments as described in Section \ref{sec:results}, but initialized K-Planes, MPPCA, and HeMPPCAT using K-Means++. Figure~\ref{fig:F_synth_supp} compares the average $\F_j$ estimation errors across the 25 datasets against $v_1$. Again, MPPCA and HeMPPCAT had similar estimation errors at lower values of $v_1$, but HeMPPCAT had much lower errors than MPPCA at larger values of $v_1$. Additionally, the non-statistical K-Planes mostly had higher errors than MPPCA and HeMPPCAT.

